\newcommand{\beq}{\begin{equation}}
\newcommand{\eeq}{\end{equation}}
\newcommand{\beqa}{\begin{eqnarray}}
\newcommand{\eeqa}{\end{eqnarray}}
\newcommand{\ba}{\begin{array}}
\newcommand{\ea}{\end{array}}

\documentclass[12pt,article]{IEEEtran}

\usepackage{a4,amsmath,epsf}
\usepackage{amssymb,bbm}
\usepackage{psfrag}
\usepackage{epsfig,epsfig,latexsym,amssymb,amsmath}
\usepackage{makeidx}
\usepackage{amsmath}
\usepackage{cite}
\usepackage{graphicx}

\newtheorem{Theorem}{Theorem}[section]

\newtheorem{Corollary}{Corollary}[section]

\newtheorem{example}{Example}[section]

\onecolumn

\renewcommand{\beq}{\beqa}
\renewcommand{\eeq}{\eeqa}

\begin{document}

\begin{titlepage}
 
\setcounter{page}{0}
 
\vskip 1.5in

\title{Polar Codes: Graph Representation and Duality}
\vskip 5.5in

\author{M. Fossorier\\
\small ETIS ENSEA/UCP/CNRS UMR-8051\\
\small 6, avenue du Ponceau,\\ 
\small 95014, Cergy Pontoise, France\\
\small Email: mfossorier@ieee.org}

\vskip 2.0in


\maketitle

\begin{abstract}
In this paper, we present an iterative construction of a polar code and develop properties
of the dual of a polar code. Based on this approach, belief propagation of a polar code
can be presented in the context of low-density parity check codes.
\end{abstract}

\thispagestyle{empty}
\end{titlepage}

\baselineskip 20pt

\section{Introduction}
\label{sec:intro}

Since their introduction~\cite{arikan:it09}, polar codes have attracted a lot of
attention due to their capacity approaching performance in near-linear time encoding and
decoding complexities ($O(N \log N)$ complexities for a polar code of length $N$). 
These theoretical results hold for very long code lengths
in conjunction with successive cancelation (SC) decoding based on the G-space representation
of the code. For a polar code of length $N$, SC decoding operates on a graph free of 4-cycles
with $N \log N$ variable nodes (associated with information bits) and $N \log N$ constraint 
nodes (associated with encoded bits).

In~\cite{yedidia:allerton02} an approach to eliminate 4-cycles in any bipartite graph associated 
with the parity check matrix of a linear code was proposed. The matrix obtained by this
approach is reminiscent of that used in SC decoding of polar codes.

In this paper, we present the graph representation of a polar code free of 4-cycle as a
generalization of the approach of~\cite{yedidia:allerton02}. We then study the dual representation
of a polar code.

\section{Graph Representation of Polar Codes: Iterative Construction}
\label{sec:graph}

\subsection{Iterative Construction for SC Decoding}
\label{sec:iter}

Let $F(m)$ represent the $m$-fold Kronecker product of the matrix $F(1)$ given by
\beqa
F(1) & = & \begin{pmatrix}1 & 0 \cr
1 & 1\cr\end{pmatrix}.
\label{eq:F1}
\eeqa
so that $F(m)$ follows the recursion
\beqa
F(m+1) & = & \begin{pmatrix}F(m) & 0 \cr
F(m) & F(m)\cr\end{pmatrix}.
\label{eq:recFm}
\eeqa
We readily observe that 4-cycles are usually present in $F(m+1)$ given in~(\ref{eq:recFm}) due to the repetition
of $F(m)$ in the second row.

A polar code of length $N=2^m$ and dimension $K$ is defined by the generator matrix obtained by selecting $K$ out of the $N$ rows of 
$F(m)$. The corresponding graph is composed of $K$ variable nodes representing the information bits and $N$ constraint
nodes representing the encoded bits. This graph can be viewed as the bipartite graph of a low density generator matrix (LDGM)
code. Equivalently, the $N-K$ deleted rows can be associated with deterministic bits (taken as 0 without loss of generality
and referred to as ``frozen bits'').
Hence all polar codes of length $N=2^m$ can be represented by the bipartite graph associated with $F(m)$.

To perform iterative decoding, this graph has to be expanded into a graph with $m \; 2^m$ variable nodes and $m \; 2^m$ constraint 
nodes~\cite{arikan:it09}. The main goal of this expansion is to remove all 4-cycles associated with the bipartite graph of $F(m)$ and 
in~\cite{eslami:it12},
it is shown that this expanded graph has girth $g=12$. This basic decomposition is guided by the way SC decoding operates and can be
viewed as a generalization of the decomposition of~\cite{schnabl:it95} for Reed-Muller (RM) codes. 
Instead, we propose to generalize
the approach developed in~\cite{yedidia:allerton02} to eliminate 4-cycles in any bipartite graph associated with the parity
check matrix of a low density parity check (LDPC) code to the case of the generator matrix of a LDGM code. 

\subsection{Review and Generalization of the Approach of~\cite{yedidia:allerton02}}
\label{sec:allerton}

For a 4-cycle associated with 2 check nodes $c_1$ and $c_2$, each containing the modulo-2 sum $x_1 + x_2$ of variable nodes
$x_1$ and $x_2$, the approach~\cite{yedidia:allerton02} introduces the new unobservable variable node $x_{12} = x_1 + x_2$ and
the corresponding constraint $c_{12} = x_1 + x_2 + x_{12} = 0$. As a result, the matrix
\beqa
H & = & \begin{pmatrix}1 & 1 \cr
1 & 1\cr\end{pmatrix}.
\label{eq:H1}
\eeqa
with column-1 and column-2 associated to variable nodes $x_1$ and $x_2$, row-1 and row-2 associated to check nodes $c_1$ and $c_2$,
is extended into the matrix
\beqa
H_e & = & \begin{pmatrix}0 & 0 & 1 \cr
0 & 0 & 1\cr
1 & 1 & 1\cr\end{pmatrix}.
\label{eq:He}
\eeqa
where the new column-3 and row-3 are associated with variable node $x_{12}$ and check node $c_{12}$, respectively.

In G-space, the previous approach can be associated with a 4-cycle associated with encoded bits $x_1$ and $x_2$, each
containing the modulo-2 sum $v_1 + v_2$ of information bits $v_1$ and $v_2$. This 4-cycle can be eliminated by
introducing the new intermediare node $v_{12} = v_1 + v_2$ and the corresponding encoded bit $x_{12} = v_{12}
+ v_1 + v_2 = 0$. This time, the matrix
\beqa
G & = & \begin{pmatrix}1 & 1 \cr
1 & 1\cr\end{pmatrix}.
\label{eq:G1}
\eeqa
with column-1 and column-2 associated to encoded bits $x_1$ and $x_2$, row-1 and row-2 associated to information bits $v_1$ and $v_2$,
is extended into the matrix
\beqa
G_e & = & \begin{pmatrix}0 & 0 & 1 \cr
0 & 0 & 1\cr
1 & 1 & 1\cr\end{pmatrix}.
\label{eq:Ge}
\eeqa
where the new column-3 and row-3 are associated with encoded bit $x_{12}$ and intermediare bit $v_{12}$, respectively.
 
\subsection{A New Iterative Construction Free of 4-Cycles}
\label{sec:iter_new}

From the method presented in Section~\ref{sec:allerton}, we expand $F(m)$ given in~(\ref{eq:recFm}) as
\beqa
F_e(m) & = & \begin{pmatrix}0 & 0 & I_{m-1}\cr
0 & F(m-1) & I_{m-1}\cr
F(m-1) & 0 & I_{m-1}\cr\end{pmatrix},
\label{eq:recFem}
\eeqa
where $I_m$ represents the $2^m \times 2^m$ identity matrix.
Based on~(\ref{eq:recFem}), we obtain the following result:
\begin{Theorem}
The matrix $F_e(m)$ can be associated with a bipartite graph with $(m+1) \; 2^{m-1}$ variable nodes and $(m+1) \; 2^{m-1}$ constraint nodes,
and girth $g=8$.
\label{theo:Fem}
\end{Theorem}
{\bf{Proof:}} 
In~(\ref{eq:recFem}), the two matrices $F(m-1)$ can be further expanded into the corresponding $F_e(m-1)$ and this operation
can be repeated iteratively until $F(1)$ is reached.
If $K_e(m,i)$ represent the size of $F_e(m)$ obtained at step-$i$ of this expansion for $1 \leq i \leq m-1$, we have
\beqa
K_e(m,1) & = & 2^{m-1} + 2 \cdot 2^{m-1} \nonumber \\
         & = & 2^{m-1} + 2^m \nonumber \\
K_e(m,2) & = & 2^{m-1} + 2 \cdot \left(2^{m-2} + 2 \cdot 2^{m-2} \right) \nonumber \\
         & = & 2 \cdot 2^{m-1} + 2^m \nonumber \\
&  \cdots & \nonumber \\
K_e(m,i) & = & i \cdot 2^{m-1} + 2^m \nonumber \\
& \cdots & \nonumber \\
K_e(m,m-1) & = & (m-1) \cdot 2^{m-1} + 2^m \nonumber \\
& = & (m+1) 2^{m-1}.
\eeqa
Finally $g>6$ directly follows from the structure of~(\ref{eq:recFem}) as it is readily seen that neither 4-cycles, nor 6-cycles
can be found in the final expansion of $F_e(m)$ at step-$(m-1)$. 

\begin{figure} [h] 
\begin{center}
\includegraphics[angle=0,scale=0.5]{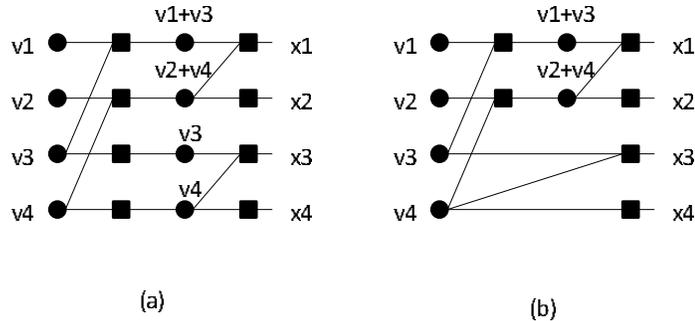}
\caption{Graph representation of $F(2)$: (a) conventional representation associated with $8 \times 8$ adjacency matrix; 
(b) proposed representation associated with $6 \times 6$ adjacency matrix.}
\label{fig:F4}
\end{center}
\end{figure}

\begin{example}
For $m=2$, the conventional graph used in SC decoding is depicted in Figure~\ref{fig:F4}-(a).
From~(\ref{eq:recFem}), we have
\beqa
F_e(2) & = & \begin{pmatrix}0 & 0 & I_1\cr
0 & F(1) & I_1\cr
F(1) & 0 & I_1\cr\end{pmatrix} \nonumber \\
& = & \begin{pmatrix}0 & 0 & 0 & 0 & 1 & 0\cr
0 & 0 & 0 & 0 & 0 & 1\cr
0 & 0 & 1 & 0 & 1 & 0\cr
0 & 0 & 1 & 1 & 0 & 1\cr
1 & 0 & 0 & 0 & 1 & 0\cr
1 & 1 & 0 & 0 & 0 & 1\cr\end{pmatrix}
\label{eq:Fe2}
\eeqa
The corresponding graph is given in Figure~\ref{fig:F4}-(b). We observe that the last two rows of $F_e(2)$
in~(\ref{eq:Fe2}) are associated with the intermediare bits $v_{13} = v_1 + v_3$ and $v_{24} = v_2 + v_4$.
\label{ex:F2}
\end{example}

The conventional graph of Section~\ref{sec:iter} developed for SC decoding can be derived from the recursion
\beqa
F_e^{SC}(m) & = & \begin{pmatrix}0 & 0 & I_{m-1} & 0\cr
0 & 0 & I_{m-1} & I_{m-1}\cr
0 & F(m-1) & 0 & I_{m-1}\cr
F(m-1) & 0 & I_{m-1} & 0\cr\end{pmatrix}.
\label{eq:recFem2}
\eeqa
However the last column of~(\ref{eq:recFem2}) can be viewed as unnecessarily redundant as it simply implies the repetition of the 
same information bits.
It follows that the conventional graph developed for SC decoding can be derived from the graph given in Theorem~\ref{theo:Fem}
with the introduction of trivial nodes only, so that we obtain a balanced graph with $m$ sections of $2^m$ variable nodes 
each, in alternance with $m$ sections of $2^m$ check nodes each (see Figure~\ref{fig:F4}). Consequently the corresponding
increase of the girth from 8 to 12
can be viewed as somewhat artificial since there is no limit in increasing the girth this way. However this insertion
of trivial nodes modifies the scheduling of iterative decoding operating on the graph and the importance of proper scheduling 
to achieve a good error performance has been discussed in~\cite{hassani:it09}. 

\section{Duality Property and Implications}
\label{sec:dual}

\subsection{Structural Properties and Graph Representation in H-space}
\label{sec:dual_prop}

Since $F(1)^2 = I_1$ and from~(\ref{eq:recFm}) 
\beqa
F(m+1)^2 & = & \begin{pmatrix}F(m)^2 & 0 \cr
0 & F(m)^2\cr\end{pmatrix},
\label{eq:recFm2}
\eeqa
a straighforward recursion provides the following theorem:
\begin{Theorem}
\beqa
F(m)^2 = I_m,
\label{eq:square}
\eeqa
\label{theo:square}
\end{Theorem}
Based on~(\ref{eq:square}), it follows:
\begin{Corollary}
Any row-$i$ of $F(m)$ is orthogonal to all rows of $F(m)^T$, but row-$i$.
\label{cor:I2}
\end{Corollary}
Furthermore, $F(m)^T$ follows the recursion:
\beqa
F(m+1)^T & = & \begin{pmatrix}F(m)^T & F(m)^T \cr
0 & F(m)^T\cr\end{pmatrix},
\label{eq:recFmT}
\eeqa
so that by recursion from $F(1)^T$, we obtain:
\begin{Corollary}
\beqa
f_{i,j}^T = f_{N-1-i,N-1-j}.
\label{eq:fT}
\eeqa
\label{cor:fT}
\end{Corollary}
Assume the rows of $F(m)$ are labeled from 0 to $2^{m}-1$ and define $S_F$ as the set of labels associated with the
frozen bits of a polar code of length $2^m$. Corollary~\ref{cor:I2} implies that the dual of this polar code
is simply defined from the rows of $F(m)^T$ in $S_F$. 
Furthermore, Corollary~\ref{cor:fT} implies that the $i$-th row of the parity check matrix $F(m)^T$ of a polar code is obtained 
from the $N-1-i$-th row of $F(m)$ written in reverse order. Hence this representation is also the representation of another polar
code of length $2^m$. 
In other words, to define a polar code $C$ and its dual $C^{\perp}$,
if a row of $F(m)$ is associated with an information bit of $C$, the same row in $F(m)^T$ is associated with a frozen bit
of $C^{\perp}$, and vice-versa. It follows that the graph representation of the parity
check matrix of a polar code is obtained from that in G-space by first reversing the representation of variable nodes and 
constraint nodes, and second inversing active and frozen bits, which also partially follows from
the duality property of a factor graph in general~\cite{forney:it01}.
This structural property between a polar code and its dual has been at least partially
observed in some works~\cite{hassani:it09}, but 
its use in iterative decoding based on belief propagation (BP) in H-space seems to have not been explicitly reported.

\begin{figure} [h]
\begin{center}
\includegraphics[angle=0,scale=0.5]{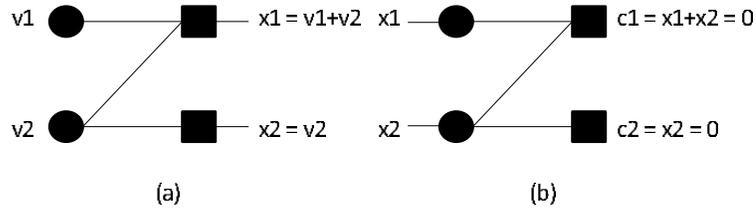}
\caption{Graph representation of: (a) $F(1)$ and (b) its dual $H(1)$.}
\label{fig:G1}
\end{center}
\end{figure}

\begin{example}
In Figures~\ref{fig:G1}-(a) and~\ref{fig:G1}-(b), the graph representations of $F(1)$ given in~(\ref{eq:F1}) and
its dual entity $H(1)$ are depicted, with
\beqa
H(1) = F(1)^T & = & \begin{pmatrix}1 & 1 \cr
0 & 1\cr\end{pmatrix}.
\label{eq:H_1}
\eeqa
\end{example}

From~(\ref{eq:recFem}), we have
\beqa
F_e(m)^T & = & \begin{pmatrix}0 & 0 & F(m-1)^T\cr
0 & F(m-1)^T & 0\cr
I_{m-1} & I_{m-1} & I_{m-1}\cr\end{pmatrix},
\label{eq:recFemT}
\eeqa
which corresponds to the expansion method of~\cite{yedidia:allerton02} in H-space applied to~(\ref{eq:recFmT}).
Consequently the results of Section~\ref{sec:iter_new} also apply directly to the dual code of a polar code
and its representation in H-space. From this graph representation free of 4-cycles, BP decoding can be performed 
as in~\cite{yedidia:allerton02} for LDPC codes.

\begin{figure} [h]
\begin{center}
\includegraphics[angle=0,scale=0.5]{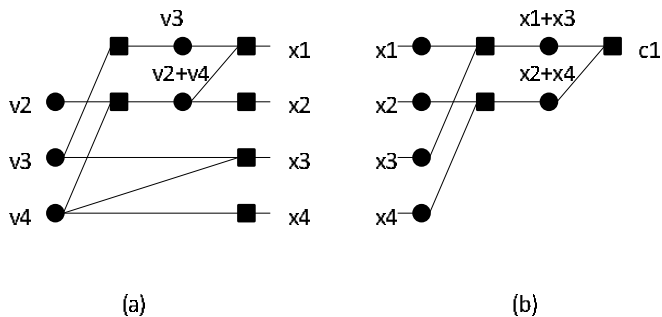}
\caption{Graph representation of RM(4,3): (a) G-space; (b) H-space.}
\label{fig:RM4_3}
\end{center}
\end{figure}

\begin{example}
Consider the (4,3) RM code obtained by freezing the information bit associated with the first row of $F(2)$.
Considering Figure~\ref{fig:F4}-(b), the graph representing this code in G-space is depicted in Figure~\ref{fig:RM4_3}-(a)
while the corresponding graph in H-space corresponding to~(\ref{eq:recFemT}) is given in Figure~\ref{fig:RM4_3}-(b).
\label{ex:RM4_3}
\end{example}

\subsection{Systematic Encoding}
\label{sec:dual_enc}

Another important consequence of these results is the fact that the systematic encoding in $O(N \log N$) computational 
complexity only suggested as ``another method'' at the end of ~\cite[Section~III-A]{arikan:cl11}
becomes straightforward and direct from the H-space representation. In the context of multistage (iterative) decoding, 
systematic encoding is not only  desirable from an implementation viewpoint, but it also minimizes the overall bit error rate 
(BER)~\cite{fossorier:it98}. To this end, we define after possible row permutation $\pi()$
\beqa
G & = & \pi \left( F(m) \right) = \begin{pmatrix} G_F \cr
G_U \cr\end{pmatrix},
\label{eq:GF}
\eeqa
with $G_F$ and $G_U$ corresponding to the frozen and unfrozen positions, respectively.
It follows that
\beqa
G^T & = & \pi \left( F(m)^T \right) = \begin{pmatrix} H_U \cr
H_F \cr\end{pmatrix},
\label{eq:HF}
\eeqa
with $G_U H_U^T = 0$. Then systematic encoding is performed by solving $[x_F x_U] H_U^T = 0$ with $x_U = v_U$
denoting the information sequence. Note that the permutation $\pi()$ necessary to obtain the form in~(\ref{eq:GF})
from the original $F(m)$ needs also to be applied to $[x_F x_U]$.

\begin{example}
Consider the (4,3) RM code defined by Figure~\ref{fig:RM4_3}-(b). Setting $x_F = x_1$ and $x_U = [x_2 x_3 x_4]
= [v_1 v_2 v_3]$ directly provides $x_1 = v_2 + v_3 + v_4$.
\label{ex:RM4_3sys}
\end{example}

This method performs systematic encoding of RM codes as a special case.

\section{Conclusion}
\label{sec:conclusion}
In this paper, we have presented the application of the concept of~\cite{yedidia:allerton02} to obtain a graph
representation of a polar code free of 4-cycles. This representation was linked to the conventional one
corresponding to SC decoding. We then investigated duality properties of polar codes so that BP decoding as
commonly used for the decoding of LDPC codes can be applied to the decoding of polar codes in H-space.

The obtention of the conventional graph from the new graph representation proposed in this paper was achieved
by the introduction of trivial nodes that artificially increases the girth of the graph from 8 to 12, but
importantly also allows to modify the scheduling of decoding based on the graph representation.
This observation suggests that scheduling can be combined with the approach of~\cite{yedidia:allerton02} to optimize error
performance. Consequently, the combination of scheduling and parallel decoding referred to as ``iterative decoding with replica''
in~\cite{zhang:it07} could be considered in the context of decoding of polar codes.

In~\cite{mori:it12}, non binary polar codes have been investigated. The generalization of the binary case
to the non binary one can be performed in several ways. We observe that the codes proposed in~\cite[Section~VIII]{mori:it12}
do not satisfy all structural properties presented in this paper for the binary case. This suggest a more
stringent definition of non binary polar codes for which all structural properties derived in this paper
could be extended to the non binary case.

\section*{Acknowledgement}
The author wishes to thank Dr. Kai Chen and Kai Niu for valuable help.

 \end{document}